**Information science and technology as applications of the physics of signalling.**

A. P. Young[1].

**General Introduction.**
The paper proposes that constraints, on structure in signalled data, are inherent in signalling mechanisms and therefore apply to all information represented by physical conditions; computers, controlled by software, are particular examples of such signalling mechanisms. The paper adopts the scientific method, proposing a theoretical model which explains the observed properties of information and of real time systems much as atomic theory explains the observed properties of matter. By proposing a model to describe the interface between the logical domain and the physical it approaches software engineering at the most fundamental level, through the physics of data generation; current thinking, by contrast, conserves and processes information only after it has been generated.

No report of similar work has been found in the literature; for this reason it is not possible to set the proposals in that context. Data flow methods (Dennis, [1]) are of particular note as they apply one feature of the model proposed. Other references are given as background reading: Shannon [2] treats the physics of signal detection in noise, Liu [3] gives a broad picture of current real time technology, Schach [4] supplements this picture with emphasis on object-oriented methods and Hoare [5] with emphasis on cooperative communication between concurrent processing threads.

The difficulties so widely experienced in complex software projects result, in the author's submission, primarily from failure to identify the physics-based model now proposed; the key components and issues have gone unrecognised, the approach to the life-cycle improvised and incoherent. If, as the author believes, constraints on logical structure are imposed by physical mechanisms they can be studied only through the physics.

**Introduction to the proposals.**
Within any signalling system physical signals communicate information from a sender to any receiver or receivers according to a convention relating physical behaviour to information, the convention established in advance. The assignment of labels to the keys of a keyboard is a simple example of such a convention, essential to connect the physical domain to the logical, a convention adopted by designers and which must be employed by users. Designers of hardware and of software implement signalling conventions by providing equipment which will apply the required physical tests as needed. In a signalling system information is generated only by the application of physical tests, to physical sources of information, thereby to perform inferences; the paper will propose a model in which the logical components, facts and inferences, have physical counterparts which are physical phenomena occupying space and time. Tests are, in this sense, the most fundamental components of the scientific model of the physical universe, the sole source of human knowledge of physical behaviour.

The paper is based on Newtonian physics.

---
[1] Email: arthur_young@realtimeart.co.uk

**Signalling.**
This section summarises long-established signalling technology and explains how it relates to the model proposed.

Signalling conventions are widely used in engineering: for example in a computer or other binary signalling system a physical indication (such as the potential difference between two terminals) of level above a given threshold might conventionally convey the information "one" while an indication of level below that threshold might conventionally convey the information "nought". A physical test, applied by any receiver, serves to distinguish the two classes of indication. In computers the sensed levels are always well above, or well below, the threshold thus ensuring that unwanted electronic noise and interference cannot cause errors of classification. Designers may adopt, concurrently, different conventions for different purposes: thus a word, stored in a computer, might convey a binary number according to one convention and the temperature within a furnace according to another, higher-level, convention. At the physical level knowledge can be acquired only by applying physical tests to classify physical signals; software provides a way of applying physical tests according to a specified procedure.

Similar methods apply where physical measurements are concerned: thus the temperature of a furnace, expressed according to a convention, identifies a class of physical phenomenon – a region, within the furnace, in which the temperature lies within a range identified by the measuring equipment employed, the test identified by that temperature succeeding.

This basic technology is now extended: natural languages, such as English or French, are also seen to provide conventions by which physical phenomena are linked to information. Thus the word " soulier", in French, is equivalent to the word "shoe" in English; a test, applied by French-speakers and by English-speakers alike, is allocated different labels in the two languages. Clearly some form of test is invoked to classify physical phenomena since one shoe may differ from another but some physical phenomena are not shoes. We might equally have invented a third language in which one class of physical phenomenon stands for another; thus a drawing of a shoe might stand for a shoe in a record of our possessions. Of course we do not understand how tests, for physical phenomena such as shoes, are performed, nor do different people necessarily apply identical tests; these reservations show only that human testing employs obscure mechanisms which vary somewhat from person to person.

While a physical system may be said to have a specific "state" at a particular point in time such states are not accessible to direct observation; any transfer of information must occupy a finite period of time as the thermal noise power, inherent in physical communications, is proportional to bandwidth and thus inversely proportional to the duration of that period (Shannon, [2]). Moreover it is clear that terms such as "shoe" identify physical phenomena which have finite life-times. For these reasons the proposed model is based on physical phenomena which occupy space-time; it is also clear that a logical model, capable of representing physical behaviour occupying space-time, must refer to components which also occupy space-time.

**The logical/physical interface.**
These arguments provide the starting-point for the model of the logical/ physical interface now proposed: a physical phenomenon is of a class *f* if, and only if, it has caused a physical test *f* to succeed. It is then termed a signal *f*. A person's knowledge of his or her physical environment originates only from physical tests applied by the sensory organs of that person; some of these tests detect the success of tests applied in physical systems within that environment. Knowledge, represented within a machine, of the nature of a physical phenomenon is contained only in the identity of a successful test or tests. According to the model proposed it can be gained only by inheritance, or by direct observation, or by inference based on existing knowledge; the precision of this knowledge depends entirely on the precision of the tests identified. (Note: the human model of physical behaviour clearly includes recognition of phenomena which are wholly or partly unobserved by the human, phenomena classified by an imagined observer. This paper is concerned only with machines).

A test is itself a signal. A person may assign, to a test applied on a particular occasion, a generic class *f'* and a specific class *f''*, *f* now equivalent to the ordered pair *(f', f'')* where *f'* identifies tests of a particular class irrespective of their outcome, *f''* identifying the outcome of a specific application of such a test. Thus a test of the generic class *weight_in_grams* might yield the result 2 when applied on a specific occasion, the test *(weight_in_grams, 2)* then succeeding; a test senses the classes of signals applied within it, generating signals of classes communicating the desired results; it may use sensing, measurement and computation in any combination, the class of the outcome or outcomes denoted by *f''* serving also to classify the input signals. Throughout its run-time any logical system performs a test composed of internal tests; the outcomes of tests may be sensed by further tests applied by human users or by other logical systems. Designs provide equipment which will perform tests in which the chosen generic classes reflect the requirements, specific classes reporting their outcomes and communicated between tests; tests are often software-controlled.

The class of a phenomenon may often be inferred by observing a sample of its life. Thus a person, observing a sample of the life of a shoe, uses existing knowledge to infer the existence of that shoe throughout an unknown period of time which contains the sampling-period. In ordinary life, as in electronics, signals are classified by sampling them, the class of the sample used to classify the entire signal.

A signal *f* (a physical phenomenon of a class allocated, according to the convention being used, an identifier *f*) may be described in greater detail only by stating first the classes of any signals known to be contained in it and second, where available, the time order in which these signals end. A signal *s* is said to end at the earliest time at which past physical behaviour implies subsequent success of the test *s*; only then can a signal *s* be detected in a physical system. One signal is said to contain another if and only if the space, occupied by the inner signal at any time within its life, then falls within the space then occupied by the outer. We may say, equivalently, that any space-time region wholly within the inner signal is also wholly within the outer.

Where two signals are known to overlap, each partly but not wholly contained in the other, then knowledge of the classes of the two signals is extended only by knowledge of the class of the signal contained in both of them.

**The structure imposed by signalling needs.**
A physical system, applying a test *f'* to classify a "wanted" signal, must also be equipped to apply the test at an appropriate time, a time at which the wanted signal will exist, the necessary physical preconditions satisfied: thus the weight of an object can validly be read only when the object is in position on the weighing machine; equally the components of an electronic message can be read only where and when they are being signalled. It follows that the necessary physical preconditions, when satisfied, must allow a test to proceed; in logical terms a boolean must become true before reading, of *f''* or of any part of it, can proceed. Where the signal is non-volatile (lasting indefinitely) the test may be applied to a sample taken at any time after the boolean has become true; a volatile signal must be tested while accessible, the testing actions initiated when the truth of the boolean is signalled. The time duration of the test *f'* may be chosen at will according to the model proposed.

For this reason any description of behaviour, represented by signals within a physical system, must take a particular structural form (see fig. 1) in which a single boolean, assigned the identifier *S*, signifies the start of description, its truth indicating that testing of the signals, by their readers, can validly proceed. The description extends as time passes; within it each boolean, on becoming true, makes accessible a fact which may contain further booleans (none, one or more), these initially untrue. Each may (or may not) later become true, making a new fact accessible, the structure growing in this way out of the single boolean *S*. This mirrors our experience of facts becoming known as time advances and of descriptions of the nature and time order of events.

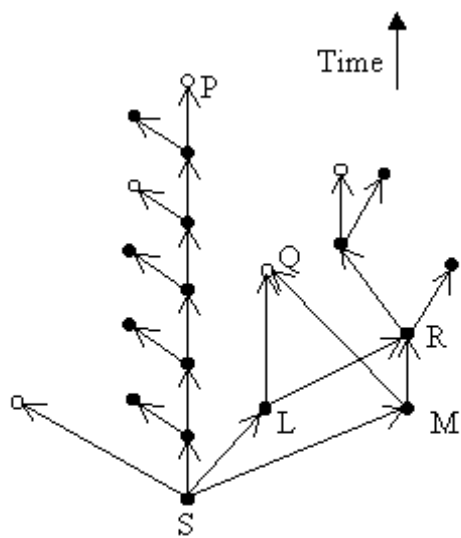

Figure 1.

Booleans are shown in fig. 1 as open circles when false and as black dots when true; arrows connect all the booleans, within a fact, to every boolean which made that fact accessible, the fact itself not shown. Within a fact the figure shows only the booleans it contains, other knowledge not represented. A single fact may be made accessible through any number of booleans thus forming a tree-like structure in which branches may recombine to form a single branch – for example to describe the behaviour of a number of physical objects which coalesce to form one.

In fig. 1 the booleans M and N both made the booleans Q and R accessible. Data flow methods (Dennis, [1]), are seen as stemming from this property of data structure.

The proposal is that information, communicated by physical signals, will be readable only if this structural form applies. Values or parts of such a structure may of course be transmitted elsewhere, time order of transmission chosen freely. Where "the latest entry" of a list is to be read directly or indirectly (a necessity in real time systems, as will be explained) booleans, giving access to those entries, may remain invisible to software. Some facts within such structures may exist only temporarily, their values discarded as garbage or lost through volatility.

**Paths.**
If each boolean, in a value of such a structure, is assigned a unique identifier then any identifier will be present in at least one list of identifiers beginning with the identifier *S* and in which each entry identifies a boolean which, on becoming true, allowed the next to be accessed, the list time-ordered. The booleans, identified by such a list, are said to form a "path" through the structure. Thus the list *[S, a, b, c, ..., q]* identifies a path in which boolean *S* gives access to a fact containing boolean *a*, this giving access to a fact containing a boolean *b*, the chain continuing until the boolean *q* is reached; this boolean may remain untrue or may give access to a fact containing no booleans. The same structure might contain other paths such as *[S, a, v, w, c, ..., q]*. In fig. 1 a list identifying, in correct order, the booleans connecting S to P would identify a path.

A fact may contain any number of booleans and may become accessible through any number of booleans; these becoming true at various times. Within such a structure the potential existence of each boolean must have been recognised in the signalling convention which applies, the convention identifying first the physical test which will, when it succeeds, indicate that the given boolean has become true and second the reading mechanism to be initiated at that time. The time order, in which booleans denoted in a path become true, is represented by the order of listing of their identifiers; the entry (such as *q* above) currently at the "latest end" of a path is termed its "latest entry". Any two identifiers will occur in the same order in all paths containing them both, the structural form limited only by this constraint and by the need for a single "earliest boolean" S.

**Inference in physical systems.**
A notation is defined to be a set in which each member is a description, a data structure of the form which has been described, the set containing every description permitted, according to the chosen method of description, to be generated within a signal of some given class *S*. A notation defines a signalling convention, a method of describing behaviour within such signals. A description is a fact and is itself composed of facts.

The truth set, of a signal *S* described using a notation *N*, is the subset of the notation *N* which includes every description capable of resulting from such an experiment. According to the model proposed knowledge of such truth sets provides the sole means of reasoning about the course of physical behaviour; we can specify a space-time region only by reference to its class, and a method of description only by reference to a notation; no other method appears possible. Knowledge of truth sets may be gained by experience or by inheritance or by some combination of the two.

Since the domain of inference, defined by S and N together, may be chosen at will it is possible to use observed behaviour to reason about behaviour inaccessible to direct observation.

Knowledge of probabilities may be gained from an experiment in which another experiment is performed repeatedly. Inferences are performed only within experiments; a test is an experiment.

According to this model a human observer may learn from observation that the use, of a sample of a signal to infer the longer-term existence of that signal, yields results consistent with human experience.

In fig.2 the large ellipse, bounded in a continuous line, represents a test performed on input signals represented by four smaller ellipses represented in the same way; three other small ellipses surrounding black areas represent signals generated as outputs of this test. Other ellipses, bounded by dotted lines, generate these input signals and sense these output signals. Tests provide the sole means of performing inferences, input and output signals representing data according to the convention which applies. In the figure area is used to represent space-time, the area representing a signal containing the areas representing signals it contains.

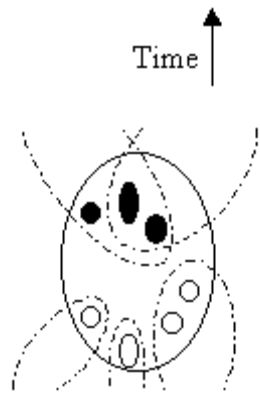

Figure 2.

Where a list of successive states describes the behaviour of an isolated physical system there is an equivalent representation in which the outcomes, of a succession of tests each performed in the system, are listed in time order, the output of each test providing the input signalled to its successor during some small period of time.

**Real time systems.**
It is suggested that real time systems are characterised by their ability to provide a service throughout a run-time of unlimited duration, a property which distinguishes them from calculators; this requires inferences of at least one generic class to be repeated without limit, a calculator defined to lack that capability. Unlimited run-time, using software of finite size, can be achieved only by repetition of at least one inference of a specific generic class. Such a repetition will generate a list in which each entry describes the outcome of one such inference, the order of listing reflecting the time order of the inferences, a topic treated more fully below. In those structures of the form described above which are generated by real time systems, these lists are dominant features; their entries are facts which may contain booleans giving access to other facts or to other lists.

In considering real time system we take the described signal $S$ to exist in the space-time region which contains the real time system and its controlled environment (the physical process it controls) throughout a period of continuous run-time. The controlled environment may contain a time-varying population of signals of which some may also contain time-varying populations of signals, a hierarchy developing in this way. Such hierarchies dictate the structural design of real time systems, a topic now addressed.

We take as an example a trader who opens and closes stores, the population of stores changing as time advances; while trading each store will take in stock and sell stock, the population of each kind of traded product, held in each store, also changing as time advances. The adopted notation might provide a list in which each new entry identifies a newly-opened store, the entry containing a boolean which is set true if and when the store closes and which gives access to any further data descriptive of the closure. An entry might also contain, for each kind of product stocked at the store in question, a list in which each entry relates to a transaction involving such a product, the list giving the history relevant to that kind of product. In this way a hierarchy is created in which the outmost signal $S$ is described by lists each describing variations in a population of signals of a particular class; each member of such a population may itself contain a time-varying population of signals, described using similar lists, the hierarchy continuing until all time-varying populations have been described. Within such a list particular entries may be annotated, or may be by-passed, when the related signal has left the population described by the list.

The hierarchy having been established in this way further lists may be required to describe changing physical conditions within a signal, recording histories of sets of such physical variables as temperature, pressure, recorded price, availability and stock-level. Thus each entry, relating to a member of a time-varying population, may provide access to a list or lists describing the history of its physical characteristics.

Figure 3 illustrates a hierarchic system: the list SP shows booleans each indicating that a fact, reporting the existence of a specific member of a first population, is accessible; of these facts all but one already give access to a further fact descriptive of that member. The typical member A contains a second population described similarly by a second list; the history of physical properties of member B typical of this population is further described by yet another list.

Structural design will, if effective, inevitably take this form, a form dictated by the physics; object-oriented methods [4], currently in vogue, appear to reflect some intuitive perception of the physical constraint.

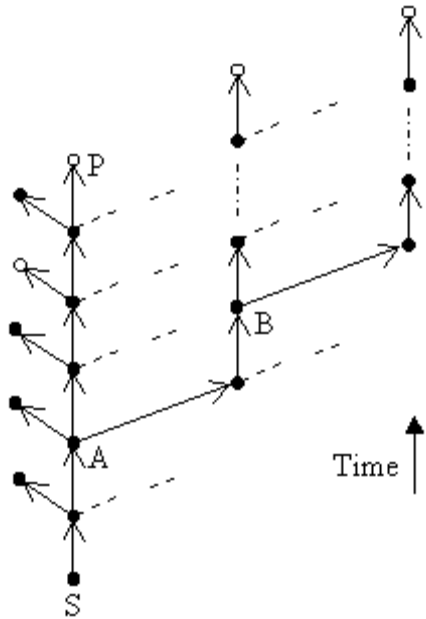

Figure 3.

It is of course permissible to employ multiple hierarchies in describing a given signal, ensuring that consistency of description is maintained. Consistency, between two potentially contradictory sources of data, requires them to be accessed through a single entry of a list: thus the behaviour of a passenger in boarding an elevator might be reported in two lists, one giving the history of the elevator and one that of the floor; a reader, reading from both lists, might learn from one that boarding had already occurred and from another that it was not yet reported. Where consistency is required values of these lists must be accessed through an entry of a single list which identifies consistent values.

It is possible to implement the proposed structure in various ways; for example a fixed population of lists might be provided, sufficient in number to accommodate the greatest-possible need, the current role of each list recorded in that list. Equally lists may be combined, reducing the number – a measure which may increase the constraints on concurrency, the topic now addressed.

"The latest entry" of any list must of course be made accessible in a way which avoids the need to enter the list at its beginning; a potentially vast list might otherwise need to be searched, from its beginning, for its latest entry. To avoid this some part of the latest entry must be stored in a location known at design-time: in current designs it is usual to store the entire entry in a location fixed for each list, each new entry overwriting its predecessor; it is envisaged that in applying the theory only the address, of the body of the latest entry, would be stored in the fixed location, to be overwritten when a new entry arrives. All lists would employ this structure, hierarchy implemented by placing, in entries of higher-level lists, references to the indirect addresses of the latest entries of lower-level lists. The latest entry of a list may then be identified and read at any time, entries retained in memory until no longer needed. A

"garbage recognition and recycling" facility is implied, as an operating-system facility, a topic treated briefly below.

The behaviour of any physical system can be described by a time-ordered list in which each entry describes the state of the system throughout a very short period of time.

**The role of data structures in inference.**
Within a logical system a value of the data structure, or of paths chosen within it, may be obtained by identifying extremities of the structure during some arbitrary period of time, an extremity marked either by a fact which contains no boolean or by a boolean which remains untrue. A value of a data structure may be extended by an inference based on relevant parts of that value. In some applications an inference will generate a fact only if the value, obtained by reading, is sufficiently extensive, performing a calculation only if its inputs have become accessible or if some other condition is satisfied. In a real time system the value of a list, returned by a reading action, will often be selected at random, dependent on the times at which booleans are accessed, the system choosing a valid value from those eligible. The truth sets of inferences must of course be chosen to provide the required behaviour; inferences must also be performed with the rapidity and urgency demanded by response-time requirements.

While any fact, generated by applying the appropriate inference to a correctly-sensed value of a data structure, will be true some facts cannot validly be used to extend the structure. Where two inferences or more, running concurrently, are able to record values of a fact accessed through a particular boolean malfunction may occur at the most basic level either because the inferences may read different values of the data structure, thus allowed to generate contradictory values of the desired fact or else, even where the facts generated are identical, because a fact made accessible by one inference may later be mutilated by another, concurrent, inference.

At higher levels malfunction due to other causes must be prevented: where the next entry of a list is to be derived from its predecessor no third entry may be allowed to separate them. Time order information, in generating one time-ordered list from another, must also be preserved. For example in a seat-reservation system a list, representing the history of a reservation for a session, must allow the writing of an entry reserving the seat only if the previous value of the list showed the seat to be unreserved; equally where an entry, in a list, reports that some event has not yet occurred and the next entry reports that it has now occurred a copy of the list, generated remotely and reversing the entry-order, would be invalid.

**Constraints on concurrency.**
The following constraints on concurrency are needed to avoid unwanted interactions occurring between inferences chosen to meet a requirement:

> Only one operation - to identify the boolean, initially false, at the latest end of a given path and to make a fact accessible through that boolean - may be in progress at any given time;

> Any fact, already used in deriving a fact accessible through a boolean in a given path must, if it remains relevant, be used in deriving any fact accessible through a boolean contained in any subsequent extension of that path.

The first restriction is necessary because inferences, concurrently generating their own values of a single fact, might otherwise conflict: for example one might use as workspace the memory occupied by a fact previously made accessible by another; or two inferences might attempt to write conflicting, although equally valid, values of the same fact since the values generated by reading will often depend on the times at which extremities of the data structure are identified. This restriction also protects logical consistency, ensuring that any fact, derived using the fact containing the boolean, will become accessible through that boolean directly rather than indirectly through an intermediate fact and boolean.

The second restriction is necessary because a system must not appear to "forget" relevant information used earlier and because time order information, contained in the data structure, must be reflected in facts generated from it. For example a list of successive values of some physical variable might be used to make a list of successive values of some function of that variable, maintaining time order information; if, through the use of concurrent computations of arbitrary time duration, the derived list becomes wrongly ordered then information has been destroyed.

**Application to structural design.**
According to the theory proposed, structural design must require identification of those classes of physical phenomenon which are significant to the application, and of which of these are contained within others of them, thus establishing the hierarchy on which requirement definition and high level design will be based. At the highest level in the hierarchy is the signal containing, throughout a period of run-time, the real time system and all the signals with which it interacts. The first task is to identify those signals, within the hierarchy, which have time-varying populations; individual signals within this hierarchy will require their own lists each describing an aspect of the history of the signal, each entry a set of sensed or measured values of physical variables. Some of these may require extension – thus an entry giving the latest position of an aircraft in Cartesian coordinates may require some extensions to give the same information in other forms and may require another extension, for example to give the distance to the nearest aircraft or airport.

The object of this phase is to identify the hierarchical structure, identifying the classes of list, of entry and of fact capable of existing within the run-time data structure. Supplemented by the use of time-stamps to indicate when booleans become true this may be used for requirement-definition, requirements expressed in a functional language such as Prolog, taking into account constraints on concurrency and defining limits on response-time performance. Specifications may apply to parts of the structure as well as to its entirety.

**Application to software design.**
High-level design has identified the classes of list which will be required at run-time; entries, within a list assigned one identity, may also, when extended in some defined way, also form a list assigned some other identity.

Functional design requires procedures to be called repeatedly to generate lists of the classes foreseen in high-level design. Procedures operate on lists named to them as parameters, appending entries or extensions according to requirement; they embody their own protection against malfunction due to concurrency, normally implemented using semaphores. The sole right, to derive and to attach a new fact to an entry of a given list, may be obtained by claiming the semaphore controlling access to that right, the semaphore released when that right has been used or relinquished. A procedure returns no data other than to indicate its own completion.

The sole right, to use a value of a first list to derive and to attach new facts to a second list or lists, may also be claimed using a semaphore; where the identities of the second lists are not known until the first list has been claimed and read (for example because each message in the first list identifies the second lists to which it is to be copied) then all potential second lists must be protected initially, the first semaphore released only when second semaphores, each protecting a second list, have all been claimed successfully.

Response-time performance is controlled independently by the mechanisms chosen to initiate procedure-calls and by the rapidity of call-execution. Changes in the content of a particular list or lists (a list of past times, perhaps, and/or a source of other input data) will mainly be used to initiate procedures, these "scheduling" lists declared with the procedures and their parameters. One procedure may also initiate others. Procedures may also be initiated by incoming "interrupt" signals and may be interrupted provided that deadlock cannot ensue.

**Memory management.**
A method of recognising "garbage" is now outlined. As run-time advances a real time system will record new facts thus extending the resident part of the data structure; if and when a fact is no longer needed it must be discarded in order to keep the size of the resident data within practical limits. One fact, the fact $S$ from which all paths within the structure originate, is resident throughout run-time and provides the only point through which the resident structure can be entered by any reader; this fact will contain information allowing facts, currently at the latest ends of lists, to be addressed. Knowledge of the address of a first fact may allow the address of a second to be determined: for example the first fact may contain the address of the second, or it may be possible to derive the address of the second from the address of the first. Such a connection, direct or indirect, between facts is termed an "access connection".

A fact can be deleted from the resident structure when two conditions have been satisfied: first all the access connections, capable of being used by new readers to obtain the address of the fact, must have been contained in facts which have themselves been deleted from the resident structure; second the fact is no longer in use, a precaution necessary to ensure that no malfunction will result from redeployment of the memory it occupies. Each fact includes information allowing this method to be applied; in the most direct approach each fact includes a semaphore to count the number of its access connections plus the number of its current users. When a semaphore reaches zero its related fact can be deleted thus destroying any access connection reliant on that fact.

A number of examples can be given:

> Where a list is at the lowest level of the hierarchy, its entries containing no further lists, then any new reader will require access only to the latest entry of the list or, exceptionally, to the latest entries conforming to some logical criterion: to the latest two entries, for example. When any new entry becomes accessible through the access connection serving the list an earlier entry will become inaccessible to any new reader employing that connection, its semaphore reduced by one. Current readers can of course continue to read from that entry until they release it by reducing its semaphore.
>
> Before a reference, from one fact to another, is created the semaphore in the referenced fact is advanced, to be reduced again when the fact, containing the reference, has been deleted.
>
> Where a fact, at a higher level within a hierarchy, reports the current existence of a signal the semaphore is advanced when the signal starts and is reduced when it ends. Any list, accessible through the fact, remains accessible to new readers only while the semaphore is non-zero.
>
> Where facts form successive entries within a list which is to be processed by multiple processors then each processor will have its own access connection (a pointer to the next entry to be processed). The semaphore, contained in a fact newly-attached to the list, will be assigned an initial value equal to the number of pointers, each processor reducing the count by one when the pointer-value, giving access to the fact, has been deleted. Although pointers are, from a conceptual viewpoint, held in lists only the latest entry of the list is ever needed, each new entry deleting its predecessor.

Clearly it will be important, in applying the methods proposed, to adopt structures in which access to semaphores is possible only through trusted software.

**Discussion.**
The scientific method has been used, proposing a simple model to explain how facts are represented and how inferences are performed in physical mechanisms. The model is consistent with experience of design at the physical level; it also explains why human knowledge of physical behaviour is selective and how reasoning, about the course of physical behaviour, can be based on knowledge thus obtained. It also explains why the behaviour of real time systems is not uniquely predictable. The model appears consistent with the widely-held belief that intelligence is an evolved capability, living things adapted to apply valid tests to their environment.

According to the theory, structural design of a real time system requires identification of the classes of signal which are significant to the application and of the hierarchy in which each of these signals may contain others; requirement definition and high level design will be based on this model. Current methods are probably damaged most by failure at this level, methods improvised intuitively without theoretical justification and without precise specification of components such as "objects".

Similar comments apply to run-time design, emphasis hitherto placed, wrongly, on the computer rather than on the underlying physics. Where inferences, such as those performed by sequential processes, run concurrently there is no fundamental need for interaction between them; each can set its own boolean, within a fact, to indicate that its results have been made accessible, a subsequent inference required to test that each boolean is set before using the associated result. If, however, a single subsequent inference is to be initiated as soon as all these booleans are true then testing must be sequential, only one test capable of succeeding; concurrency, in generating the fact that all results have become available, must be controlled as in any other fact-generating procedure. Cooperative communication, between concurrent threads, provides one way of implementing these tests within list-processing operations; more direct ways, minimising constraints on timing, appear preferable.

While the paper has concentrated on information technology the proposed theory may prove relevant to wider studies of intelligence and learning, and to physics beyond the realm of Newton.

**References.**
1. Dennis, Data Flow Supercomputers, IEEE Computer Vol. 18 No. 11, Nov. 1980, pp 48-56.
2. Shannon, C, Communication in the presence of noise, Proc. Inst. Radio Engineers Vol. 37, No.1, pp10-21, Jan. 1949.
3. Liu, JWS, *Real Time Systems.* Prentice Hall London (2003), ISBN 9780130996510.
4. Schach SR, Object-oriented and Classical Software Engineering, WCB/McGraw-Hill 2006.
5. Hoare, CAR, *Communicating Sequential Processes.* Prentice-Hall London (1985), ISBN 0-8053-153271-8.